\documentclass[pra,a4paper,superscriptaddress]{revtex4}
\usepackage{amsmath}
\usepackage{amsfonts}
\usepackage{graphicx}
\usepackage{longtable}
\newcommand{\be}{\begin{equation}}
\newcommand{\ee}{\end{equation}}

\newcommand{\ba}[1]{\left(\begin{array}{#1}}
\newcommand{\ea}{\end{array}\right)}
\begin{document}

\title{ What does monogamy in higher powers of a correlation measure mean?}
\author{P. J. Geetha}\affiliation{Department of Physics, Kuvempu University, 
Shankaraghatta, Shimoga-577 451, India}
\author{Sudha }
\affiliation{Department of Physics, Kuvempu University, 
Shankaraghatta, Shimoga-577 451, India}
\affiliation{Inspire Institute Inc., Alexandria, Virginia, 22303, USA.}
\author{A. R. Usha Devi}
\affiliation{ Department of Physics, Bangalore University, Bangalore 560 056, India}
\affiliation{Inspire Institute Inc., Alexandria, Virginia, 22303, USA.}
\date{\today}
\begin{abstract} 
We examine here the proposition that all multiparty quantum states can be made monogamous by considering  positive integral powers of any quantum correlation measure. With Rajagopal-Rendell quantum deficit as the measure of quantum correlations for symmetric $3$-qubit pure states, we illustrate that monogamy inequality is satisfied for higher powers of quantum deficit. We discuss the drawbacks of this inequality in quantification of correlations in the state.  We also prove a monogamy inequality in  higher powers of classical mutual information and bring out the fact that such inequality need not necessarily imply restricted shareability of correlations. We thus disprove the utility of higher powers of any correlation measure in establishing monogamous nature in multiparty quantum states. 
\end{abstract}

\maketitle

\section{Introduction}
\label{intro}

It is well known that classical correlations are infinitely shareable whereas there is a restriction on the shareability of quantum entanglement amongst the several parts of a multipartite quantum state~\cite{ckw,osb,ter,proof}. The concept of monogamy of entanglement and monogamy of quantum correlations has been studied quite extensively~[1--18] and it is shown that the measures of quantum correlations such as quantum discord~\cite{oz}, quantum deficit~\cite{akrqd} are not monogamous for some category of pure states~\cite{prabhu,sudha}. The polygamous nature of quantum correlations other than entanglement has initiated discussions on the properties to be satisfied by a correlation measure to be monogamous~\cite{bruss} and it is shown that a measure of correlations is in general non-monogamous if it does not vanish on the set of all separable states~\cite{bruss}. 

While it has been shown that~\cite{sqd} square of the quantum discord~\cite{oz} obeys monogamy inequality for $3$-qubit pure states,     
an attempt to show that {\emph {all multiparty states can be made monogamous}} by considering higher integral powers of a non-monogamous quantum correlation measure has been done in Ref.~\cite{salini}. It is shown that (See Theorem 1 of Ref.~\cite{salini}) if $Q$ is a non-monogamous correlation measure and is monotonically decreasing under
discarding systems then $Q^n$, $n=2,\,3,\cdots $ can be a monogamous correlation measure for tripartite states~\cite{salini}. With quantum work-deficit $Q_{wd}$ as a correlation measure, it is numerically shown that almost all $3$-qubit pure states become monogamous when fifth power of ${Q}_{wd}$ is considered~\cite{salini}. 

In this work, we analyze the implications of the proposition~\cite{salini} that higher integral powers of a quantum correlation measure reveal monogamy in all multiparty quantum states. Towards this end we first verify the above proposition by adopting quantum deficit~\cite{akrqd}, an operational measure of quantum correlations for our analysis. Quantum deficit has been shown to be, in general, a non-monogamous measure of correlations for $3$ qubit pure states~\cite{sudha}. In Ref.~\cite{sudha}, the monogamy properties (with respect to quantum deficit) of symmetric $3$-qubit pure states belonging to $2$-, $3$- distinct Majorana spinors classes~\cite{maj} has been examined and it has been shown that all states belonging to the $2$-distinct spinors class (including W-states) are polygamous. It has also been shown~\cite{sudha} that the superposition of W, obverse W states, belonging to the SLOCC class of $3$-distinct Majorana spinors~\cite{maj}, are polygamous with respect to quantum deficit. Here we consider both these classes of states and illustrate that they can be monogamous with respect to higher powers of quantum deficit. We examine the possibility of quantification of tripartite correlations using monogamy relation in higher powers of a quantum correlation measure and illustrate that such an exercise is unlikely to yield fruitful results. 
  
In order to analyze the relevance of monogamy with respect to higher integral powers of a quantum correlation measure, we bring forth a monogamy-kind-of-an-inequality in higher powers of classical mutual information~\cite{nc}. The possibility of a monogamy relation in higher powers of a classical correlation measure  even in the arena of classical probability theory raises questions regarding the meaning attributed to such an inequality.  We discuss this aspect and bring out the fact that monogamy in higher powers of a quantum correlation measure need not necessarily imply limited shareability of  correlations.  

Quantum deficit, a useful measure of quantum correlations was proposed by Rajagopal and Rendell~\cite{akrqd} while enquiring into the circumstances in which entropic methods can distinguish the quantum separability and classical correlations of a composite state. It is defined as the relative entropy~\cite{nc} of the state $\rho_{AB}$ with its classically decohered counterpart $\rho^d_{AB}$. That is,   
\begin{eqnarray}
D_{AB}&=&S(\rho_{AB}\vert\vert \rho^d_{AB}) \nonumber \\
&=&\mbox{Tr} (\rho_{AB} \ln \rho_{AB})-\mbox{Tr} (\rho_{AB} \ln \rho^d_{AB}).   
\end{eqnarray}
is the quantum deficit of the state $\rho_{AB}$ and it determines the quantum excess of correlations in $\rho_{AB}$ with reference to its classically decohered counterpart $\rho_{AB}^d$. As $\rho^d_{AB}$ is diagonal in the eigenbasis  $\{\vert a\rangle\}$, $\{\vert b\rangle\}$ of the subsystems $\rho_A$, $\rho_B$ (common to both $\rho_{AB}$, $\rho_{AB}^d$) one can readily evaluate $D_{AB}$ as~\cite{sudha} 
\begin{eqnarray}
\label{newd}
D_{AB}&=&\mbox{Tr} (\rho_{AB} \ln \rho_{AB})-\mbox{Tr} (\rho_{AB} \ln \rho^d_{AB})\nonumber \\ 
&=& \sum_i \lambda_i \ln \lambda_i-\sum_{a,b} P_{ab}\ln P_{ab}, 
\end{eqnarray}
where $\lambda_i$ are the eigenvalues of the state $\rho_{AB}$ and 
$P_{ab}=\langle a, b\vert\rho_{AB}\vert a, b\rangle$ denote the diagonal elements of $\rho_{AB}^d$.
Through an explicit evaluation of the quantum deficit $D_{AB}$, the polygamous nature (with respect to quantum deficit $D_{AB}$) of two SLOCC inequivalent classes of symmetric $3$-qubit pure states has been illustrated in Ref.~\cite{sudha}. In particular, it is shown that~\cite{sudha} the monogamy relation 
\be
\label{qdmon}
D_{AB}+D_{AC}\leq D_{A:BC}
\ee 
is {\emph{not satisfied}} for symmetric $3$-qubit states with $2$-distinct Majorana spinors~\cite{maj}. Amongst the $3$-qubit GHZ and superposition of W, obverse W states, the monogamy inequality (\ref{qdmon}) is satisfied by  the GHZ states while the superposition of W and obverse W states does not obey it~\cite{sudha} inspite of both the states belonging to the SLOCC family of $3$-distinct spinors~\cite{maj}. 

In the following we illustrate that symmetric $3$-~qubit pure states obey monogamy relation
in higher powers of quantum deficit $D_{AB}$. The states of interest are given by,  
\begin{eqnarray}
\label{w}
\vert \rm{W} \rangle &=& \frac{\vert 100 \rangle+\vert 010 \rangle+\vert 001 \rangle}{\sqrt{3}}  \nonumber \\ 
\label{ww}
\vert \rm{ W\bar{W}} \rangle &=& \frac{\vert {\rm W} \rangle +\vert {\rm{\bar W}} \rangle}{\sqrt{2}}   
\end{eqnarray}  
where $\vert {\rm{\bar W}}\rangle=\frac{\vert 011 \rangle+\vert 101 \rangle+\vert 110  \rangle}{\sqrt{3}}$ is the obverse W state. 

The reduced density matrices of the $3$ qubit W state are given by 
\begin{eqnarray}
\rho_{AB} &=& \rho_{AC}=\frac{1}{3}\ba{cccc} 1 & 0 & 0 & 0 \\ 0 & 1 & 1 & 0 \\ 0 & 1 & 1 & 0 \\0 & 0 & 0 & 0 \ea \mbox{and} \nonumber \\ 
\rho_A &=& \rho_B=\rho_C=\frac{1}{3}\ba{cc} 2 & 0 \\ 0 & 1 \ea
\end{eqnarray}  
With $\chi_1=(1,\,0)$, $\chi_2=(0,\,1)$ being the eigenvectors of $\rho_A$,
the decohered counterpart $\rho_{AB}^d$ of $\rho_{AB}$ is obtained as~\cite{sudha}
\begin{eqnarray}
\rho^d_{AB}&=&\mbox{diag}\,\left( P_{11}, P_{22},\,P_{33},\,P_{44} \right) \nonumber \\ 
&=&\mbox{diag}\,\left(\frac{1}{3},\,\frac{1}{3},\,\frac{1}{3},\,0\right)=\rho^d_{AC}; \\   P_{ii}&=&\langle \chi_i,\chi_i\vert \rho_{AB}\vert \chi_i, \chi_i\rangle; \nonumber 
\end{eqnarray}
As $\lambda_1=\frac{2}{3}$, $\lambda_2=\frac{1}{3}$ are the non-zero eigenvalues of $\rho_{AB}$, we obtain the quantum deficit $D_{AB}$ to be,  
\be
D_{AB}=\frac{2}{3}\ln \frac{2}{3}+\frac{1}{3}\ln \frac{1}{3}- \ln \frac{1}{3} \approx 0.462
\ee
An evaluation of the eigenvectors $\eta_j$, $j=1,\,2,\,3,\,4$ of the bipartite subsystems $\rho_{AB}=\rho_{AC}$ of the state $\vert {\rm W}\rangle$ allows us to find out the decohered counterpart $\rho^d_{A:BC}$ of the state  $\rho_{ABC}$ and we have 
\begin{eqnarray}
\rho^d_{A:BC}&=&\mbox{diag}\,\left(P_{11},\, P_{12},\,P_{13},\,P_{14},\,P_{21},\, P_{22},\,P_{23},\,P_{24} \right) \nonumber \\ 
&=& \mbox{diag} \, \left(0,\,0,\,\frac{2}{3},\,0,\,0,\,0,\,0,\,\frac{1}{3}\right); \\  P_{ij}&=&\langle \chi_i,\eta_j\vert \rho_{\rm W}\vert \chi_i, \eta_j\rangle, \ \ \rho_{\rm W}=\vert {\rm W} \rangle \langle {\rm W} \vert. \nonumber 
\end{eqnarray} 
The quantum deficit $D_{A:BC}$ of the W state is thus obtained as,
\be
D_{A:BC}=0-\left(\frac{2}{3}\ln \frac{2}{3}+\frac{1}{3}\ln \frac{1}{3} \right) \approx 0.636. \nonumber
\ee
It is easy to see that
\be
D_{AB}+D_{AC}=2 D_{AB}\approx 2 \times 0.462 > D_{A:BC} \approx 0.636 
\ee
and the monogamy inequality  (\ref{qdmon}) is not obeyed~\cite{sudha}.
  
For the state  $\vert \rm{ W\bar{W}}\rangle$, the reduced density matrices are
\begin{eqnarray}
\rho_{AB}&=&\rho_{AC}=\frac{1}{6}\ba{cccc} 1 & 1 & 1 & 0 \\ 1 & 2 & 2 & 1 \\ 1 & 2 & 2 & 1 \\0 & 1 & 1 & 1 \ea; \nonumber \\
\rho_A &=& \rho_B=\rho_C=\frac{1}{6}\ba{cc} 3 & 2 \\ 2 & 3 \ea
\end{eqnarray}
and their common non-zero eigenvalues are $\lambda_1=\frac{2}{3}$, $\lambda_2=\frac{1}{3}$. 
The decohered density matrices $\rho^d_{AB}$, $\rho^d_{A:BC}$ are respectively given by~\cite{sudha}
\begin{eqnarray}
\rho^d_{AB}&=&\mbox{diag}\,\left(\frac{3}{4},\,\frac{1}{12},\,\frac{1}{12},\,\frac{1}{12}\right)  \nonumber \\
& & \nonumber \\
\rho^d_{A:BC}&=&\mbox{diag} \, \left(\frac{5}{6},\,0,\,0,\,0,\,0,\,\frac{1}{6},\,0,\,0\right)
\end{eqnarray}
The quantum deficit $D_{AB}(=D_{AC})$ and $D_{A:BC}$ are therefore obtained as
\[
D_{AB}=\frac{2}{3}\ln \frac{2}{3}+\frac{1}{3}\ln \frac{1}{3}-\left(\frac{3}{4} \ln \frac{3}{4}+\frac{3}{12}\ln \frac{1}{12}\right) \approx 0.386 
\]
\be
D_{A:BC}=0-\left(\frac{5}{6}\ln \frac{5}{6}+\frac{1}{6}\ln \frac{1}{6} \right) \approx 0.45.
\ee
Here too we have
\be
D_{AB}+D_{AC}=2 D_{AB}\approx 2 \times 0.386 > D_{A:BC} \approx 0.45
\ee
and the monogamy inequality (\ref{qdmon}) is not obeyed~\cite{sudha}. 
 
Having illustrated the polygamous nature of the states $\vert {\rm W}\rangle$, $\vert {\rm W{\bar W}}\rangle$ with respect to quantum deficit, we wish to see whether higher powers of quantum deficit indicate monogamy in these and if so for what powers. In Table~I, we have tabulated $D^n_{AB}$, $D^n_{A:BC}$ and the value of $D_{A:BC}^{n}-2 D_{AB}^{n}$, $n\geq 1$ for both  $\vert {\rm W}\rangle$.    
\begin{table}[ht] 
\caption{Monogamy w.r.t integral powers of Quantum Deficit for $3$ qubit pure symmetric states}
\begin{center}
\scriptsize{\textbf{
\begin{tabular}{|c|c|c|c|c|}
\hline
  &  & \multicolumn{2}{c|}{Quantum Deficit} &  \\
	\cline{3-1} \cline{4-1}
State & Powers  &  $D_{AB}^{n}$ & $D_{A:BC}^{n}$ &  $D_{A:BC}^{n}-2 D_{AB}^{n}$ \\
 & ($n$) & ($=D_{AC}^{n}$)& & \\
\hline\hline 
  & $1$ & $0.462$ &  $0.636$ & $-0.287$ \\
	\cline{2-1} \cline{3-1} \cline{4-1} \cline{5-1} 
$3$ qubit& $2$ & $0.213$ &  $0.405$ & $-0.022$ \\
	\cline{2-1} \cline{3-1} \cline{4-1} \cline{5-1} 
 W	& $3$ & $0.098$ &  $0.257$ & $0.060$ \\
\cline{2-1} \cline{3-1} \cline{4-1} \cline{5-1} 
  & $4$ & $0.045$ &  $0.164$ & $0.072$ \\
\cline{2-1} \cline{3-1} \cline{4-1} \cline{5-1} 
	& $5$ & $0.021$ &  $0.104$ & $0.062$ \\
\hline \hline  
 & $1$ & $0.386$ &  $0.450$ & $-0.322$ \\
	\cline{2-1} \cline{3-1} \cline{4-1} \cline{5-1} 
 $3$ qubit	& $2$ & $0.149$ &  $0.203$ & $-0.095$ \\
	\cline{2-1} \cline{3-1} \cline{4-1} \cline{5-1} 
 $\rm{W\bar{W}}$ & $3$ & $0.057$ &  $0.091$ & $-0.023$ \\
\cline{2-1} \cline{3-1} \cline{4-1} \cline{5-1} 
  & $4$ & $0.022$ &  $0.041$ & $-0.003$ \\
\cline{2-1} \cline{3-1} \cline{4-1} \cline{5-1} 
	& $5$ & $0.008$ &  $0.018$ & $0.0013$ \\
\hline
\end{tabular}}}
\end{center}
\end{table}

It can be readily seen from the table that though the $3$-qubit states $\vert {\rm W} \rangle$ and 
  $\vert {\rm W{\bar W}}\rangle$ are polygamous with respect to quantum deficit, its third power satisfies monogamy inequality for $\vert {\rm W} \rangle$ whereas fifth power of quantum deficit is required for making the state $\vert {\rm W{\bar W}}\rangle$ monogamous. 
  
We have also examined the monogamy with respect to $D^n_{AB}\left(=D^n_{AC}\right)$, $D_{A:BC}$ for an arbitrary  symmetric $3$-qubit pure state belonging to the family of $2$-distinct spinors~\cite{maj}. The state is given by~\cite{sudha} 
\be
\vert \psi \rangle \equiv \cos \frac{\theta}{2} \vert 000 \rangle+ \sin \frac{\theta}{2} \left(\frac{\vert 100 \rangle+\vert 010 \rangle+\vert 001 \rangle}{\sqrt{3}}\right) 
\ee
with $0<\theta<\pi$ and for $\theta=\pi$, we get the $\vert {\rm{W}}\rangle$ state. An explicit evaluation of $D_{AB}$, $D_{A:BC}$, as a function of $\theta$, has been done in Ref. ~\cite{sudha} and the state is seen to be polygamous for all values $\theta$. But in 
higher powers of $D_{AB}$, $D_{A:BC}$, monogamy inequality is satisfied and as the power $n$ increases more states become monogamous. Fig.~1 illustrates this feature. 
\begin{figure}[ht]
\centerline{\includegraphics* [width=2.4in,keepaspectratio]{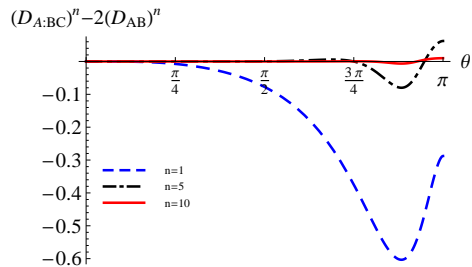}} 
\caption{The plot of $(D_{A:BC})^n-2(D_{AB})^n$ versus $\theta$ for $3$-qubit pure states with $2$-distinct spinors} 
\end{figure}

At this juncture, it would be of interest to know whether quantification of non-classical correlations in a tripartite state is possible through the monogamy inequality in higher powers of the correlation measure.  Notice that monogamy inequalities in squared concurrence~\cite{ckw}, squared negativity~\cite{neg} have been useful in quantifying the tripartite correlation through  concurrence tangle~\cite{ckw} and negativity tangle~\cite{neg}. A systematic attempt to quantify the correlations in $3$-qubit pure states using square of quantum discord as the measure  
of quantum correlations has been done in Ref.~\cite{sqd}.  We observe here that in order to quantify the tripartite correlations using monogamy inequality in $Q^n$, one has to consider the non-negative quantity $\tau_q=Q^r_{A:BC}-Q^r_{AB}-Q^r_{AC}$, where $r\geq 1$ is the {\emph{minimum degree} at which the state becomes monogamous. But there immediately arise questions regarding whether  the minimum degree $r$ of $Q$ has any bearing on the amount of non-classical tripartite correlations in the state. In fact, every correlation measure requires a different integral power $r$  to reveal monogamous nature in a $3$-qubit state~\cite{salini}. Whereas fifth power of quantum discord is sufficient to make almost all $3$-qubit pure states monogamous~\cite{salini}, we have seen here that one requires still higher powers in quantum deficit ($r\geq 10$) (See Fig. 1) for some pure symmetric $3$-qubit states. As such there does not appear to be any relation between the non-classical correlations in the state and the minimum degree $r$ of a correlation measure. 
For instance, we have (See Table~I) 
\begin{eqnarray*}
D^r_{A:BC}-D^r_{AB}-D^r_{AC}&=&0.06 \  \mbox{for } \ \vert {\rm{W}}\rangle \ \mbox{when}\ r=3 \\ 
D^r_{A:BC}-D^r_{AB}-D^r_{AC}&=&0.0013 \  \mbox{for} \  \vert {\rm{W{\bar W}}}\rangle \ \mbox{when}\ r=5.
\end{eqnarray*}
Also, as  $D_{A:BC}-D_{AB}-D_{AC}=1$ for $3$-qubit GHZ state~\cite{sudha}, we have 
$D^r_{A:BC}-D^r_{AB}-D^r_{AC}=1$ at $r=1$ for GHZ state.   
It is not apparent whether states having higher correlations possess larger $r$ with smaller value of $\tau_q$ or vice versa. 
Whichever be the case, the way in which $r$ can be accommodated in finding the tripartite correlations is not evident even in these simplest examples.  Also, the monogamy relation in higher powers of a correlation measure $Q$ will not have the physical meaning of restricted shareability of correlations if $Q^r$ is not established as a 
proper correlation measure satisfying essential properties such as local unitary invariance. The tripartite correlation measure $\tau_{q}$ should also be established as a valid correlation measure for each $r$~\footnote{For square of quantum discord $\tau_q$ is shown to satisfy all properties of a correlation measure in Ref.~\cite{sqd}.}. Without addressing these issues, a mere quantification of tripartite correlations through $\tau_q$ may not yield justifiable results. 

\section{Monogamy-kind-of relation in higher powers of mutual information}
We now go about exploring the meaning associated with monogamy in positive integral powers of a correlation measure. Towards this end, we prove  
a monogamy-kind-of-a relation in higher powers of classical mutual information and investigate its consequences.      

From the strong subadditivity property of Shannon entropy~\cite{nc}, we have, 
\be
\label{subad}
H(X,Y,Z) + H(Y) \leq H(X,Y) + H(Y,Z).
\ee
Casting Eq. (\ref{subad}) in terms of the  mutual information~\cite{nc} 
\begin{eqnarray*}
H(X:Y)&=&H(X) + H(Y) - H(X,Y), \\  
H(Y,Z)&=&H(Y) + H(Z)- H(Y:Z). 
\end{eqnarray*}
we obtain
\[
H(X,Y,Z) + H(X:Y) - H(X) \leq H(Y) + H(Z)- H(Y:Z).
\]
and this implies 
\be
\label{subad3}
H(X:Y)+ H(Y:Z) \leq H(X) + H(Y) + H(Z)-H(X,Y,Z). 
\ee
In view of the fact that 
\be
H(Y:XZ) = H(Y) + H(X,Z) - H(X,Y,Z) \nonumber 
\ee
where $H(Y:XZ)$ denotes the mutual information between the random variables $Y$, $XZ$,  
we make use of the relation $H(X,Y,Z) = H(Y) + H(X,Z) - H(Y:XZ)$ in Eq. (\ref{subad3}) to obtain 
\begin{eqnarray}
H(X:Y) + H(Y:Z) &\leq& H(X) + H(Y) + H(Z) - \left [H(Y)+ H(X,Z) - H(Y:XZ)\right] \nonumber \\ 
&\leq& H(X) + H(Z) - H(X,Z) + H(Y:XZ). \nonumber
\end{eqnarray} 
As $H(X:Z)=H(X) + H(Z) - H(X,Z)$,  $H(X:Y)=H(Y:X)$ 
we obtain the relation 
\be
\label{subadfi}
H(Y:X) + H(Y:Z) \leq H(Y:XZ) + H(X:Z)
\ee
obeyed by the trivariate joint probability distribution $P(X,Y,Z)$ indexed by the random variable $XYZ$. 
Notice that the relation 
\be
\label{mon1}
H(Y:X) + H(Y:Z) \leq H(Y:XZ)
\ee 
represents a monogamy relation between the random variables $X$, $Y$ and $Z$. But this inequality is not true due to the existence of the non-negative term $H(X:Z)$ on the right hand side of Eq. (\ref{subadfi}). That is we have, 
\be
\label{subadfi2}
H(Y:X) + H(Y:Z) \geq H(Y:XZ) 
\ee
According to Theorem~1 of Ref. \cite{salini}, a non-monogamous measure of correlations satisfies monogamy inequality in higher integral powers when the measure is decreasing under removal of a subsystem. Thus, in order to prove the monogamy relation in higher powers of mutual information, we need to establish that  
\[
H(Y:X) \leq H(Y:XZ),\ \  H(Y:Z) \leq H(Y:XZ).
\] 
We show in the following that mutual information indeed is non-increasing under removal of a random variable.        

On making use of the relations 
\begin{eqnarray} 
H(X:Y) &=& H(X) + H(Y) - H(X,Y),\nonumber \\  
H(Y:XZ) &=& H(Y) + H(X,Z) - H(X,Y,Z) \nonumber 
\end{eqnarray}
we have, 
\begin{eqnarray}
\label{mono2}
H(Y:XZ) - H(X:Y)& =& H(Y) + H(X,Z) - H(X,Y,Z) -\left[H(X) + H(Y) - H(X,Y)\right] \nonumber \\
 &=&H(X,Y)+H(X,Z)-H(X)-H(X,Y,Z)       
\end{eqnarray}
As $H(Z|XY)=H(X,Y,Z) - H(X,Y)$, $H(Z|X)=H(X,Z)-H(X)$ denote the respective conditional entropies,  Eq. (\ref{mono2}) simplifies to  
\be 
H(Y:XZ)-H(Y:X)=H(Z|X) - H(Z|XY).  \nonumber  
\ee
Using the fact that conditioning reduces entropy, i.e., $H(Z\vert XY)\leq H(Z\vert X)$, we readily have 
\be
\label{monofi1}
H(Z|X)-H(Z|XY) \geq 0 \ \mbox{or} \ H(Y:X)\leq H(Y:XZ)
\ee
One can similarly show that 
\be
\label{monofi2}
H(Y:Z)\leq H(Y:XZ).  
\ee 
We are now in a position to prove the monogamy relation  
\be
\label{hi}
H^n(Y:X) + H^n(Y:Z) \leq H^n(Y:XZ) \ \ \mbox{for} \ \ n>r
\ee
based on the proof of Theorem~1 of Ref.~\cite{salini}. Here $r>1$ is the lowest integer for which the above equality is satisfied.   

Denoting  
\be
H(Y:XZ) = x, \ \ H(Y:X) = y,\ \  H(Y:Z) = z
\ee 
we have $x < y + z$,  $x > y > 0$, $x > z > 0$ and hence 
\[
0< y/x < 1, \ \  0 < z/x  < 1
\]
which follow from the non-negativity of mutual information and from Eqs.(\ref{subadfi2}), (\ref{monofi1}), (\ref{monofi2}).   
This implies $\lim_{m\rightarrow \infty} (y/x)^{m} = 0$, $\lim_{\rightarrow \infty} (z/x)^{m} = 0$ and hence 
$\forall\  \epsilon > 0$ there exist positive integers $n_1(\epsilon), n_2(\epsilon)$ such that,
\begin{eqnarray}
& & (y/x)^{n} < \epsilon \ \ \forall \ \ \ \mbox{positive integers} \ \ n\geq m_1(\epsilon) , \nonumber \\
& & (z/x)^{n} < \epsilon \ \ \forall \ \ \ \mbox{positive integers} \ \ n\geq m_2(\epsilon) 
\end{eqnarray}
With a choice of $\epsilon = \epsilon_1 < \frac{1}{2}$ we get $0<(y/x)^{n} < \epsilon_1$ and $0<(z/x)^{n} < \epsilon_1$, $\forall$ positive integers $n\geq m(\epsilon_1)$, where $m(\epsilon_1)= \max\{m_1(\epsilon_1),m_2(\epsilon_1)\}$ we readily obtain the inequality  
\be
\label{hi2}
(y/x)^{n}+ (z/x)^{n} < 1 \ \forall \ n\geq m(\epsilon_1)
\ee 
which is essentially the monogamy relation Eq. (\ref{hi})  

Having established the monogamy relation in higher powers of classical mutual information (See Eq. (\ref{hi})), we now examine its implications. In fact, we are interested in knowing whether the monogamy relation in higher powers of a correlation measure (classical/quantum) reflects restricted shareability of correlations. Towards this end we raise the following questions.
\begin{itemize} 
\item[(a)]  Does  Eq. (\ref{hi}) imply that the distribution of bipartite correlations between 
$X$, $Y$ and $Y$, $Z$ are restrictively shared among the random variables $X$, $Y$, $Z$ in the trivariate probability distribution $P(X,\,Y,\,Z)$?  
\item[(b)] If a classical correlation measure satisfies monogamy inequality in its higher powers, does it mean limited shareability of classical correlations in a quantum state?
\item[(c)] What does the monogamy relation satisfied by higher power of a non-monogamous measure of quantum correlations mean?     
\end{itemize} 
An affirmative answer to (a) and (b) negates the unrestricted shareability of classical correlations.  
But it is well known that classical correlations in a multiparty quantum state can be distributed at will amongst its parties. 
This implies we need to negate both the statements (a) and (b). Now it is not difficult to see that negation of (a) and (b) immediately provides an answer to (c):  \\
{\emph {Existence of a monogamy relation in higher powers of any correlation measure (classical or quantum) does not necessarily mean restricted shareability of correlations in a multiparty state}}. 

\section{Conclusion}
In conclusion, we have illustrated that monogamy relation satisfied in higher powers of a non-monogamous correlation measure is not useful either to quantify the correlations or to signify that all mutiparty states have restricted shareability of correlations. We hope that this work is helpful in clarifying whether or not higher powers of quantum correlation measure are to be taken up for examining the monogamous nature of quantum states.

\section*{Acknowledgements}
P. J. Geetha  acknowledges the support of Department of Science and Technology (DST), Govt. of India through the award of INSPIRE fellowship.

\end{document}